\documentclass[12pt]{iopart}
\pdfoutput=1

\usepackage{iopams}  
\usepackage[mathscr]{eucal}
\usepackage{graphicx}

\usepackage{hyperref}

\usepackage{cite}

\begin{document}

\title[Living on the edge of instability]{Living on the edge of instability}

\author{Artem Ryabov$^{1,2,*}$, Viktor Holubec$^{1,3}$,\\ and Ekaterina Berestneva$^{1}$}

\address{$^1$ Charles University, Faculty of Mathematics and Physics, Department of Macromolecular Physics, V~Hole{\v s}ovi{\v c}k{\' a}ch 2, CZ-180~00 Prague, Czech Republic}

\address{$^2$ Centro de F{\' i}sica Te{\' o}rica e Computacional, Departamento de F{\' i}sica, Faculdade de Ci{\^ e}ncias, Universidade de Lisboa, Campo Grande P-1749-016 Lisboa, Portugal}

\address{$^3$ Universit{\"a}t Leipzig, Institut f{\"u}r Theoretische Physik, Postfach 100 920, D-04009 Leipzig, Germany}

\ead{rjabov.a@gmail.com}

\vspace{10pt}
\begin{indented}
\item[] 09.08.2019 
\end{indented}


\begin{abstract}
Statistical description of stochastic dynamics in highly unstable potentials is strongly affected by properties of divergent trajectories, that quickly leave  meta-stable regions of the potential landscape and never return. Using ideas from theory of Q-processes and quasi-stationary distributions, we analyze position statistics of non-diverging trajectories. We discuss two limit distributions which can be considered as (formal) generalizations of the Gibbs canonical distribution to highly unstable systems. Even though the associated effective potentials differ only slightly, properties of the two distributions are fundamentally different for all highly unstable system. The distribution for trajectories conditioned to diverge in an infinitely distant future is localized and light-tailed. The other distribution, describing trajectories surviving in the meta-stable region at the instant of conditioning, is heavy-tailed. The exponent of the corresponding power-law tail is determined by the leading divergent term of the unstable potential. We discuss different equivalent forms of the two distributions and derive properties of the effective statistical force arising in the ensemble of non-diverging trajectories after the Doob h-transform. The obtained explicit results generically apply to non-linear dynamical models with meta-stable states and fast kinetic transitions.
\end{abstract}

\vspace{2pc}
\maketitle


\section{Introduction}

Unstable stochastic dynamics is ubiquitous in non-linear models. Its characteristic feature is fast divergence of trajectories that have left a meta-stable state. 
The divergence restricts precision and duration of experiments  \cite{Siler/etal/SciRep2017, Siler/etal/PRL2018}, and limits applicability of the standard statistical analysis based on the averages, because the latter rapidly diverge with time \cite{Filip/Zemanek/JOpt2016, Zemanek/etal/JOpt2016, Ornigotti/etal/PRE2018}.

In recent theoretical and experimental works 
\cite{Ornigotti/etal/PRE2018, Siler/etal/PRL2018}, the standard description based on moments has been replaced by analysis of local characteristics of the system given by the mode and the curvature near the maximum of the probability density function. The second important point was the shift of attention from the complete ensemble of trajectories to the conditional statistics of non-divergent ones. 
This allowed to obtained a detailed picture of the unstable dynamics. The theory has been applied to the paradigmatic case of the one-dimensional overdamped  Brownian motion 
\begin{equation}
\label{eq:Langevin}
 \frac{{\rm d} X_t }{{\rm d} t} = -\frac{1}{\gamma} V'\!\left(X_t\right) + \sqrt{2 D } \xi_t,
\end{equation}
in the highly unstable cubic potential
\begin{equation}
V(x) = \frac{\mu }{3} x^{3}.
\end{equation}
Above, the standard Gaussian white noise satisfies $\langle \xi_t \rangle =0$, $\langle \xi_t \xi_{t'} \rangle = \delta(t-t')$, and the diffusion coefficient $D$  is proportional to the thermal energy $k_{\rm B}T$ in accordance with the Einstein relation $D=k_{\rm B}T/\gamma$, where $\gamma$ stands for the friction constant. 

The main focus of the aforementioned works has been on the short-time dynamics, zero-noise limit and the so-called quasi-stationary distribution in highly unstable systems. In the present work, we use ideas from the probabilistic theory of conditioned stochastic processes~\cite{Chetrite/Touchette/AHP2015, bookQSD} and derive generic behavior of non-diverging trajectories. 

First, in Sec.~\ref{sec:highlyunstable}, we define precisely what we understand by highly unstable dynamics, demonstrate typical features of trajectories, and emphasize that their divergence is related to the loss of normalization of the propagator. In Sec.~\ref{sec:Schrodinger}, we show that highly unstable systems have a discrete spectrum and prove an identity for the generalized partition function. The latter allows us to derive tails of the both discussed limit distributions (Sec.~\ref{sec:PItails} and Sec.~\ref{sec:QSD}), and the magnitude of an effective statistical force arising in the conditioned ensemble from the Doob h-transform~\cite{DoobBook} (\ref{sec:Fs}). 

The quasi-stationary distribution (Sec.~\ref{sec:QSD}) and the limit distribution of the trajectories that never diverge (Sec.~\ref{sec:PIst}) can be regarded as generalizations of the Gibbs canonical distribution to appropriate ensembles of highly unstable systems. Both these distributions reduce to the Gibbs canonical one for stable potentials. Analysis of their properties generic for highly unstable dynamics, particularly based on the notion of effective potentials, are the main topic of the present work. 

Alternative methods that can be successfully applied to study highly unstable dynamics include analysis of first-passage times \cite{RednerBook, HanggiRevModPhys1990, Arecchi1982, YoungPRA1985, SanchoPRA1989, HirschPRA1982, Sigeti1989, SanchoPRA1991, Caceres1995, MantegnaPRL1996, AgudovPRE1998, Lindner2003, Brunel2003, FiasconaroPRE2005, CaceresJSP2008, Ryabov/etal/PRE2016, Siler/etal/SciRep2017}, and of the so-called nonlinear relaxation times \cite{AgudovMalakhovPRE1999, DubkovPRE2004}. The analysis of {\em times} yields e.g.\ transition rates for unstable processes.  
Our present approach is complementary to these methods in the sense that it provides a detailed picture of the particle {\em position} statistics in different naturally arising conditional ensembles. 

\section{Highly unstable dynamics}
\label{sec:highlyunstable}

By {\em highly unstable} dynamics we shall understand an overdamped Brownian motion in a potential $V(x)$ decreasing towards $-\infty$ for large $|x|$ so fast that the particle {\em reaches (minus) infinity in a finite time}. We shall call such potentials as highly unstable. 
Equivalently, they can be defined as the class of potentials for which the mean first-passage time from any point $x$ to $x=\pm \infty$ is finite \cite{Ryabov/etal/PRE2016}. 

The simplest examples of highly unstable potentials include the cubic, and the inverted quartic potentials. Generally speaking, highly unstable potentials decrease towards minus infinity at least as $-|x|^n$ for large $|x|$, with $n>2$. 
To see that such potentials are highly unstable, we inspect the zero-noise limit of the Langevin equation~(\ref{eq:Langevin}) with $V(x)\sim -|x|^n$ when $x \to -\infty$. 
For the initial condition $X_0=y$ placed into the unstable region, $y<0$, we obtain the deterministic solution 
\begin{equation}
X_t = - \left[ \frac{1}{|y|^{2-n} - (n-2) \mu t  }\right]^{\frac{1}{n-2} },
 \qquad n>2.  
 \label{eq:Xtzeronoise}
\end{equation} 
 This solution reaches minus infinity at the finite time $t_\infty=|y|^{2-n}/[(n-2)\mu]$, when the denominator attains zero. 
Notice that for the inverted parabolic potential ($n=2$), we get the exponential divergence of $X_t$. Even though the exponential divergence can be considered as ``fast'' in some contexts, it is much ``slower'' than the divergence generated by Eq.~(\ref{eq:Xtzeronoise}). Therefore, the inverted parabolic potential is not highly unstable. 
  
Equivalently, one can determine the condition $n>2$ starting from the exact expression for the mean-first passage time to (minus) infinity. The mean time is known analytically in one dimension \cite{GardinerSM}. For the detailed discussion related to the high instability of the cubic potential see e.g.\ Ref.~\cite{Ryabov/etal/PRE2016} and references therein. 

\begin{figure}[t]
\includegraphics[width=0.99\textwidth]{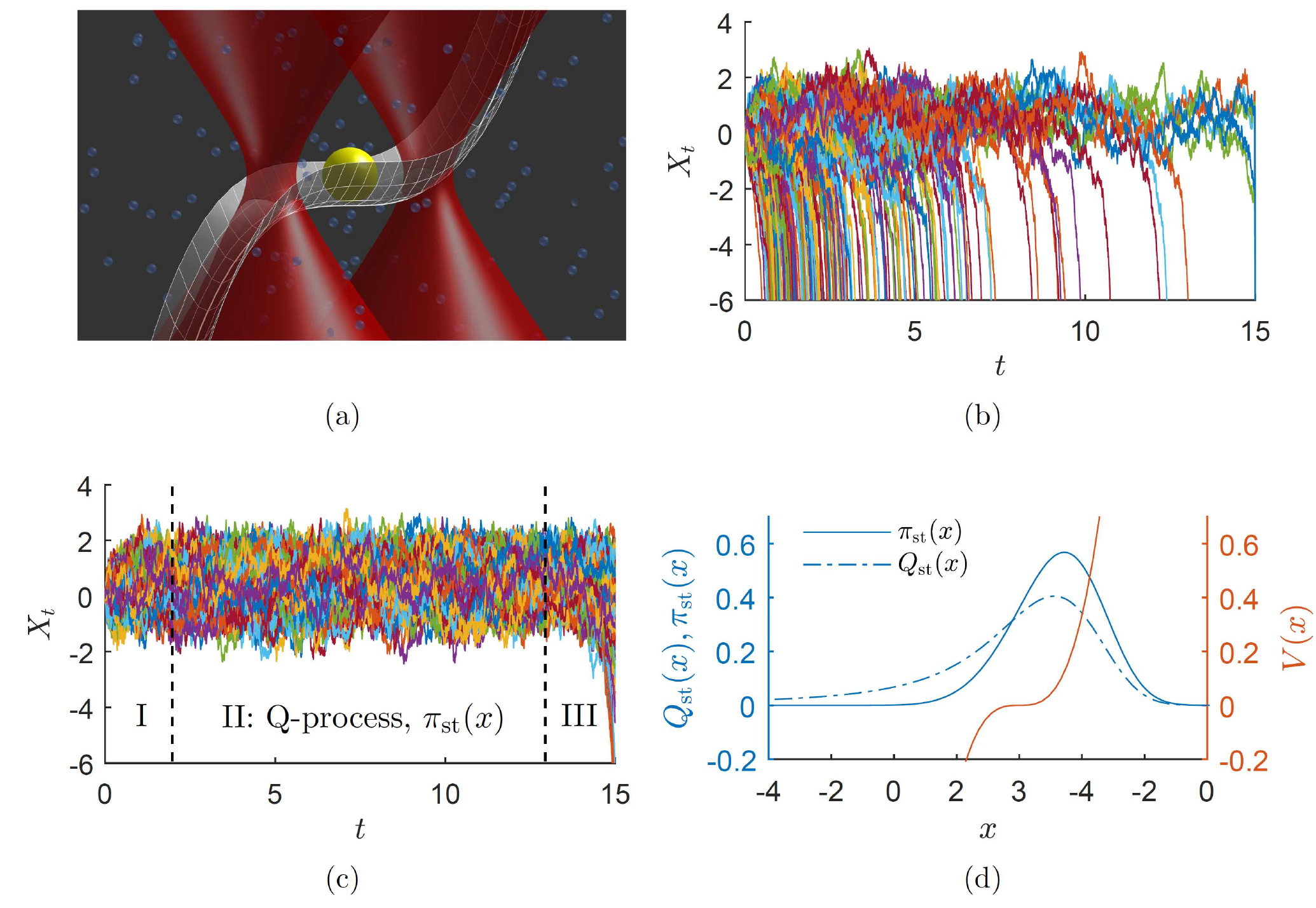}
     \caption{(a)~Sketch of the Brownian particle in the highly unstable cubic optical potential discussed in Refs.~\cite{Siler/etal/SciRep2017, Siler/etal/PRL2018}. 
     (b)~Characteristic feature of highly unstable dynamics are rapidly diverging  trajectories that never return to finite $x$. 
     (c)~Ensemble of trajectories that do not diverge at least up to time $t=15$. Their dynamics exhibits two transient (I and III) and a stationary regime (II). 
     (d)~Right axis: The cubic potential used to generate trajectories for panels (b) and (c). 
     (d)~Left axis: The light-tailed limit distribution $\pi_{\rm st}(x)$, Eq.~(\ref{eq:PIstVeff}), describes statistics in the stationary regime II in the panel (c). The heavy-tailed quasi-stationary distribution $Q_{\rm st}(x)$, Eq.~(\ref{eq:Qsteff}), corresponds to non-diverging trajectories at the last instant of regime III, i.e., to $t=15$ in the panel~(c). } 
     \label{fig:FIG1}
\end{figure}

Figure~\ref{fig:FIG1}(b) shows a typical sample of trajectories starting  at $y=0$ and diffusing in the highly unstable cubic potential. 
A trajectory in the sample can consist of three distinct segments. 
First, on the potential plateau, i.e., in the region near the origin approximately determined by the condition that the cubic potential is weak compared to thermal energy, $|V(x)|/k_{\rm B}T < 1$,  the trajectory undergoes the (almost) free diffusion. 
Second, occasionally, the trajectory is reflected by the strong cubic repulsion and returns quickly to the plateau if it reaches a large positive $x$ [upward spikes in Fig.~\ref{fig:FIG1}(b)].
Third, the last part of the trajectory begins if the particle moves left from the plateau and reaches highly unstable region $x<0$. There, the decreasing branch of the cubic potential is extremely strong compared to the thermal noise, $|V(x)|/k_{\rm B}T \gg 1$, and the trajectory is quickly dragged towards minus infinity. The divergence is nearly deterministic, in accord with Eq.~(\ref{eq:Xtzeronoise}) for $n=3$. 

A characteristic feature of highly unstable dynamics, which we prove in Sec.~\ref{sec:Schrodinger}, is that the Fokker-Planck operator corresponding to the Langevin equation~(\ref{eq:Langevin}), 
\begin{equation} 
\hat{\mathcal{L}}^{\dagger} = D \partial^{2}_{xx} + \gamma^{-1}\, \partial_{x} V'(x), 
\end{equation}
has a discrete spectrum of eigenvalues
\begin{equation}
\label{eq:lambdaineq}
0 < \lambda_0 < \lambda_1 < \ldots ,
\end{equation}
\begin{equation}
\label{eq:eigenvaluepn}
\hat{\mathcal{L}}^{\dagger} p_n(x) = - \lambda_n p_n(x). 
\end{equation}
That is, for highly unstable dynamics, there exists an isolated leading decay rate $\lambda_0$, and the strictly positive gap 
\begin{equation}
\Delta= \lambda_1 - \lambda_0, 
\end{equation}
between the eigenvalues corresponding to the slowest and the second-slowest decaying modes. 

The propagator, or the probability density function (PDF) of the particle position at time $t$, given that initially the particle was at $y$, satisfies the Fokker-Planck equation
\begin{equation}
\partial_t P(x,t|y) = \hat{\mathcal{L}}^{\dagger} P(x,t|y).
\end{equation} 
The eigenvalues determine decay rates of individual modes with amplitudes given by the eigenfunctions $p_n(x)$ defined in Eq.~(\ref{eq:eigenvaluepn}). The propagator can be expressed in terms of the eigenfunctions as \cite{HakenSynergetics} 
\begin{equation} 
\label{eq:propagator}
P(x,t|y) = \sum_{n=0}^{\infty} p_n(x) s_n(y) {\rm e}^{-\lambda_n t}.
\end{equation}
Above, the functions $s_n(y)$ depend on the initial particle position, $y$, and follow from the adjoint eigenvalue problem
\begin{equation}
\hat{\mathcal{L}}\, s_n(y) = - \lambda_n s_n(y),
\end{equation}
with the operator 
\begin{equation}
\hat{\mathcal{L}} = D \partial^{2}_{yy} - \gamma^{-1} V'(y) \partial_{y}.
\end{equation} 
adjoint to the Fokker-Planck operator $\hat{\mathcal{L}}^\dagger$. The adjoint operator is called the generator of the Markov process. 
 The two sets of eigenfunctions satisfy the following normalization conditions
\begin{eqnarray}
\label{eq:normp0n}
& \int & p_n(x) {\rm d}x = 1, \\  
\label{eq:normpsn}
& \int & s_m(x) p_n(x) {\rm d}x = \delta_{m,n}.
\end{eqnarray}
We shall refer to $p_n(x)$ and $s_n(y)$ as the right and left eigenfunctions of the Fokker-Planck operator $\hat{\mathcal{L}}^\dagger$, respectively.

The high instability of dynamics manifests itself through exponential decay of normalization $S(t|y)$ of the propagator $P(x,t|y)$~(\ref{eq:propagator}).  
The normalization equals to the probability that a trajectory has not diverged up to time $t$ and thus we call it as the survival probability \cite{RednerBook}. In symbols, we have 
\begin{equation}
S(t|y)={\rm Prob}\{\tau_{\rm d}>t | X_0=y \} = \int P(x,t|y) {\rm d}x . 
\end{equation} 
Above, $\tau_d$ stands for the time of divergence of a trajectory. 
According to Eq.~(\ref{eq:propagator}), in the long-time limit only the most-slowly decaying term contributes significantly to the sum, i.e.,  
\begin{equation}
P(x,t|y) = p_0(x) s_0(y) {\rm e}^{-\lambda_0 t} \left[1+O({\rm e}^{-\Delta t}) \right], 
\end{equation}
which we will simply write as
\begin{equation}
\label{eq:propagatorasy}
P(x,t|y) \sim  p_0(x) s_0(y) {\rm e}^{-\lambda_0 t} , 
\end{equation}
and 
\begin{equation}
\label{eq:Sasy}
S(t|y) \sim  s_0(y) {\rm e}^{-\lambda_0 t},
\end{equation}
where ``$\sim$'' means that the ratio of the two expressions converges to unity with increasing time. 

Physical meaning of the exponential decay~(\ref{eq:Sasy}) can be understood based on the illustration in Fig.~\ref{fig:FIG1}(b). The decay arises from the fast divergence of majority of the trajectories which reach $X_t=-\infty$ in a finite time. They are not included in the statistics described by the propagator $P(x,t|y)$ for any finite $x$, $x\in (-\infty, +\infty)$, and their weight is given by $[1-S(t|y)]$.  

Finally, notice that for a confining potential $V_{\rm c}(x)$, the dynamics becomes stable and conservative, and we have $\lambda_0=0$, $s_0(x)=1$, and $s_0(x)p_0(x)=p_0(x)=\pi_{\rm eq}(x) = {\rm e}^{-\beta V_{\rm c}(x)}/Z$, where the equilibrium partition function ensures the normalization condition, $Z=\int {\rm e}^{-\beta V_{\rm c}(x)} {\rm d}x$. Contrary to this, in the highly unstable non-conservative case, the eigenfunction $s_0(x)$ is no-longer constant. Below, we shall see that the expressions $Q_{\rm st}(x)=p_0(x)$ and  $\pi_{\rm st}(x)=s_0(x)p_0(x)$, which are identical in the stable case, play roles of stationary distributions in proper conditioned ensembles of non-diverging trajectories. The two ensembles, described by these distributions, have fundamentally different statistical properties, cf.\ Sec.~\ref{sec:PItails} and Sec.~\ref{sec:QSD}. 

\section{Schr{\" o}dinger equation: Normalization of eigenvectors}
\label{sec:Schrodinger}

The eigenproblem for the non-Hermitian Fokker-Planck operator involves distinct left and right eigenfunctions for each eigenvalue $-\lambda_n$. The two sets of eigenfunctions satisfy different boundary conditions and should be normalized properly to form orthogonal spaces, as we have described above in Sec.~\ref{sec:highlyunstable}. 

\begin{figure}[t]
    \centering
\includegraphics[width=0.99\textwidth]{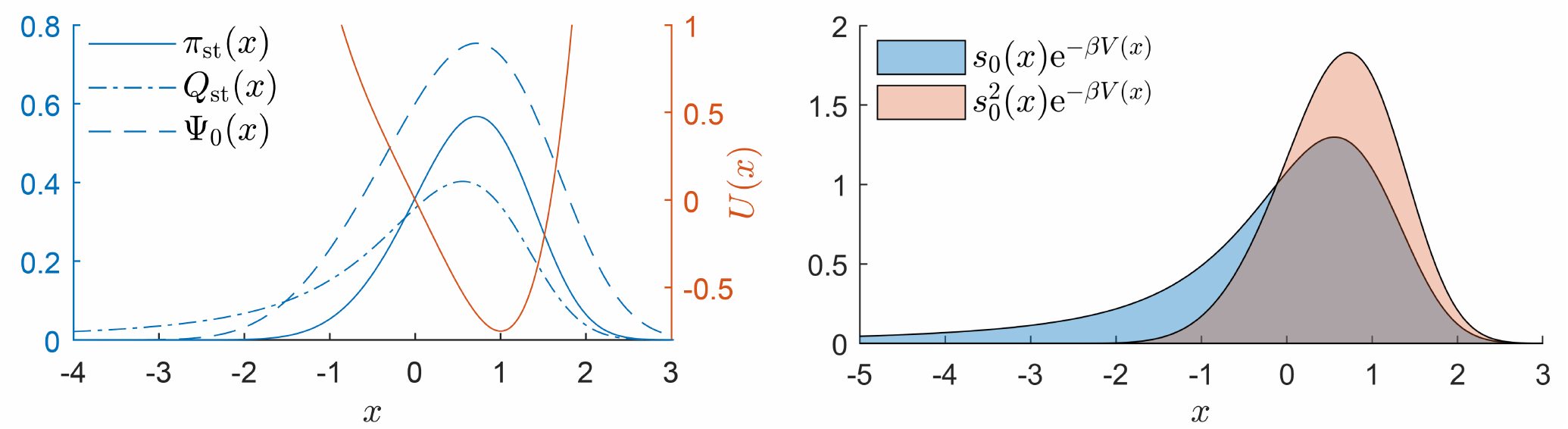}
     \caption{Left: The ground state wave function (dashed line) in the potential $U(x)$, compared with the limit distribution of the Q-process~(\ref{eq:PIstp0s0}) (dot-dashed line) and the quasi-stationary distribution~(\ref{eq:Qsts0}) (solid line). The function are related through Eqs.~(\ref{eq:PSIp0}) and (\ref{eq:PSIs0}). 
Right: the two functions occurring under integrals in the identity~(\ref{eq:identityZ}) enclose the same areas. Up to the normalization constant $Z$, the heavy-tailed function ${\rm e}^{-\beta V(x)} s_0(x)$ equals $Q_{\rm st}(x)$, and the localized function ${\rm e}^{-\beta V(x)} s_0^2(x)$ equals $\pi_{\rm st}(x)$, see Eqs.~(\ref{eq:Qsts0}) and (\ref{eq:PIsts02}), respectively.} 
     \label{fig:psi}
\end{figure}

The non-Hermitian eigenproblem can be transformed into the regular Hermitian one by a similarity transformation that recasts the Fokker-Planck operator (or equivalently the generator) into the Hermitian Hamiltonian  \cite{RiskenFPE, GardinerSM, VanKampenSP}:
\numparts
\begin{eqnarray} 
\hat{\mathcal{H}} &=& {\rm e}^{\beta V/2}\, \mathcal{L}^\dagger\, {\rm e}^{-\beta V/2},  \\
\hat{\mathcal{H}} &=& {\rm e}^{-\beta V/2}\, \mathcal{L}\, {\rm e}^{\beta V/2},  \\
\hat{\mathcal{H}} &=& - D \partial_{xx}^2 + U(x),
\end{eqnarray}  
\endnumparts 
where $\beta=1/k_{\rm B}T$, and the diffusion coefficient, $D$, determines the inverse mass of the representative quantum particle. The transformed potential, 
\begin{equation} 
U(x) = \frac{1}{4D}\left[ \frac{V'(x)}{\gamma} \right]^2\! - \frac{V''(x)}{2\gamma}, 
\end{equation} 
is confining for all highly unstable potentials $V(x)$. 
For the cubic potential, $U(x)$ is a quartic well [the term proportional to $(V'(x))^2$] with a slight asymmetry and shift of the minimum caused by the linear term arising from $V''(x)$, see Fig.~\ref{fig:psi}. 

The transformation provides two valuable insights. The first results from the discrete structure of the spectrum of $\hat{\mathcal{H}}$ that is equivalent to the discrete well-separated spectrum of $\hat{\mathcal{L}}^\dagger$. Indeed, the spectrum of $\hat{\mathcal{H}}$ is identical to that of $ \hat{\mathcal{L}}^\dagger$ up to the sign: 
\begin{equation} 
\hat{\mathcal{H}}\, \Psi_n(x) = \lambda_n \Psi_n(x). 
\end{equation} 
Because $U(x)$ is always confining for highly unstable $V(x)$, the spectrum is always discrete as it is the case for all confining infinite potential wells \cite{Berezin/Shubin/SE}. Therefore, the inequalities~(\ref{eq:lambdaineq}) follow directly from the high instability of $V(x)$.

The second insight arises from the transformation between the wave functions $\Psi_n(x)$ and eigenfunctions of $\hat{\mathcal{L}}^\dagger$. 
The ground state wave function, associated with the energy $\lambda_0$, is related to the eigenfunctions $p_0(x)$ and $s_0(x)$ of $\hat{\mathcal{L}}^\dagger$ via  
\begin{eqnarray}
\label{eq:PSIp0}
& & p_0(x) = \frac{1}{\sqrt{Z}}\, {\rm e}^{-\beta V(x)/2}\, \Psi_0(x),\\
\label{eq:PSIs0}
& & s_0(x) = \sqrt{Z}\, {\rm e}^{\beta V(x)/2}\, \Psi_0(x),
\end{eqnarray}
respectively. The factor $Z$ is independent of $x$ and it plays no role in the construction of the propagator~(\ref{eq:propagator}) for the diffusion problem. 
Hence, $Z$ is usually omitted in standard literature, e.g.\ \cite{RiskenFPE, GardinerSM, VanKampenSP}, when discussing the spectral solution of the Fokker-Planck equation. 
On the contrary, the factor $Z$, to which we shall refer as to {\em the generalized partition function}, is fundamental for the understanding of the Q-process. 

Two equivalent integral representations of the partition function, used below to explore properties of stationary distributions, are
\numparts 
\begin{eqnarray}
\label{eq:Zs}
Z &=& \int {\rm e}^{-\beta V(x)} s_0(x) {\rm d} x,\\
\label{eq:Zs2}
Z &=& \int {\rm e}^{-\beta V(x)} s_0^2(x) {\rm d} x.
\end{eqnarray}
\endnumparts 
Hence, we obtain the identity  
\begin{equation}
\label{eq:identityZ}
\int {\rm e}^{-\beta V(x)} s_0(x) {\rm d} x =
\int {\rm e}^{-\beta V(x)} s_0^2(x) {\rm d} x .
\end{equation} 
Even though the functions ${\rm e}^{-\beta V(x)} s_0(x)$ and ${\rm e}^{-\beta V(x)} s_0^2(x)$ are strikingly different, they enclose equal areas. Figure~\ref{fig:psi} shows both the localized function  
 ${\rm e}^{-\beta V(x)} s_0^2(x)$ with light tails, and the function ${\rm e}^{-\beta V(x)} s_0(x)$ possessing a heavy tail for negative $x$. 

The identity~(\ref{eq:identityZ}) follows  from the normalization conditions~(\ref{eq:normp0n}) and (\ref{eq:normpsn}), which, for the eigenfunctions $p_0(x)$ and $s_0(x)$, read 
\begin{eqnarray}
\label{eq:normp0}
& \int & p_0(x) {\rm d}x = 1, \\ 
& \int & s_0(x) p_0(x) {\rm d}x = 1.
\label{eq:normps}
\end{eqnarray}
 The combination of Eqs.~(\ref{eq:PSIp0}) and (\ref{eq:PSIs0}) yields
\begin{eqnarray}
\label{eq:p0usings0}
& & p_0(x) = \frac{1}{Z} \, {\rm e}^{-\beta V(x)}\, s_0(x), \\ 
& & s_0(x) p_0(x) = \frac{1}{Z} \, {\rm e}^{-\beta V(x)}\, s_0^2(x).
\label{eq:PIstUsings0}
\end{eqnarray} 
The requirement that right-hand sides of both equations are normalized to unity immediately gives us the identity~(\ref{eq:identityZ}). 
In other words, the constant $Z$ in Eqs.~(\ref{eq:PSIp0}) and (\ref{eq:PSIs0}) is crucial for consistency of Eqs.~(\ref{eq:p0usings0}) and (\ref{eq:PIstUsings0}), relating the eigenfunctions $p_0(x)$ and $s_0(x)$, with the normalization conditions~(\ref{eq:normp0}) and (\ref{eq:normps}). 

For the Gibbs canonical equilibrium distribution in a confining potential $V_{\rm c}(x)$, the generalized partition function $Z$ is identical to the equilibrium partition function. In equilibrium, we have $\lambda_0=0$, $s_0(x) = 1$, and the corresponding right eigenfunction of $\hat{\mathcal{L}}^\dagger$ is the Gibbs canonical distribution $p_0(x)={\rm e}^{-\beta V_{\rm c}(x)}/Z$, with $Z=\int {\rm e}^{-\beta V_{\rm c}(x)}{\rm d}x$. 

\section{Q-process: Dynamics}

Figure~\ref{fig:FIG1}(c) shows three distinct dynamical regimes observed on trajectories that do not diverge in a highly unstable potential $V(x)$. In the transient regime I, the relaxation of the initial condition takes place. In the stationary regime II, the initial condition is already forgotten and the ensemble of trajectories resembles sample paths of a stationary conservative process diffusing away from the instability region. Within II, the particle position is described by the time-independent PDF $\pi_{\rm st}(x)$. Finally, in the transient regime III, the distribution departs from $\pi_{\rm st}(x)$ and approaches the heavy-tailed quasi-stationary distribution $Q_{\rm st}(x)$ discussed in Sec.~\ref{sec:QSD}. 

Durations of the transient regimes I and III can be estimated by the inverse difference of decay rates $1/\Delta$, $\Delta=\lambda_1-\lambda_0$, that is finite due to high instability of $V(x)$. The finite transient times allow for an exact statistical description of trajectories in the regimes I and II. 
Formally, this is done by shifting the last transient regime III towards infinite times. This limit has no significant influence on trajectory statistics within regimes I and II. Remarkably, the limit transforms the non-Markovian process, conditioned on a non-local property (non-divergence up to a given time), into a Markovian one, that can be analyzed using standard tools as the Langevin and the Fokker-Planck equation. 

Let us choose an ``intermediate'' time $t$, and the ``final'' time of conditioning $\tau$, such that $0 \leq t \leq \tau$. In Fig.~\ref{fig:FIG1}(c), we have $\tau=15$ and $t\in [0,15]$. All trajectories of interest are assumed to diverge after time $\tau$. That is, the time of divergence $\tau_{\rm d}$ of any trajectory satisfies $\tau_{\rm d}>\tau$. Now, we can formally shift the conditioning time $\tau$ to infinity and define the distribution of trajectories at finite time $t$:
\begin{equation} 
\label{eq:pidef}
\pi(x,t|y) = \lim_{\tau \to \infty } 
 \left< \delta\left( X_t - x \right) \right>_{ \tau_{\rm d} > \tau} 
= \lim_{\tau \to \infty } \frac{\int P(z,\tau; x,t|y) {\rm d}z}{S(\tau |y)}.
\end{equation}
The limit $\tau \to \infty$ means that the process is conditioned to never diverge. In mathematical literature, such a process is commonly denoted as the Q-process \cite{Lambert/EJP2007, Meleard/Villemonais/PS2012}. 

The propagator $\pi(x,t|y)$ for the Q-process is derived by a direct evaluation of the above limit. Employing the Markov property, we rewrite the joint probability density in~(\ref{eq:pidef}) as the product of propagators for the intervals $[0,t]$ and $[t,\tau]$:   
$P(z,\tau; x,t|y)=P(z,\tau-t|x)P(x,t|y)$. 
Then, for large $\tau$ and fixed $t$,  we get 
\begin{equation}
\label{eq:Qderivation}
\fl \qquad 
 \frac{\int P(z,\tau-t|x)P(x,t|y)  {\rm d}z}{S(\tau |y)}  \sim 
 \frac{\int p_0(z) s_0(x){\rm e}^{-\lambda_0 (\tau-t)} P(x,t|y)  {\rm d}z}{s_0(y){\rm e}^{-\lambda_0 \tau} }, 
 \quad \tau \to \infty. 
\end{equation} 
The conditioning time $\tau$ cancels out and the integral over $z$ equals to one. The limit shifts the time of conditioning and the whole transient region III in Fig.~\ref{fig:FIG1}(c) towards infinite times and removes information on III from the dynamics of trajectories at any fixed $t$. 

The right-hand side of Eq.~(\ref{eq:Qderivation}) yields explicit form of the propagator for the Q-process:
 \begin{eqnarray} 
\label{eq:pi}
\pi(x,t|y) = \frac{s_0(x)}{s_0(y)}\, {\rm e}^{\lambda_0 t} P(x,t|y).
\end{eqnarray} 
It is straightforward to verify that the Q-process is a conservative Markov process. The normalization condition, 
$\int \pi(x,t|y) {\rm d}x =1$, 
is verified most easily using the eigenfunction expansion of $P(x,t|y)$~(\ref{eq:propagator}) and the orthogonality relation for the eigenfunctions~(\ref{eq:normpsn}). The Chapman-Kolmogorov equation, 
$\pi(x,t_1+t_2|y)=\int \pi(x,t_2|z) \pi(z,t_1|y) {\rm d}z$, follows directly from the Chapmann-Kolmogorov equation for the bare propagator $P(x,t|y)$. 

The Q-process is the driven diffusion process and, as such, its propagator $\pi(x,t|y)$ satisfies the backward equation 
\begin{equation} 
\partial_t \pi(x,t|y) = \hat{\mathcal{L}}_{\rm s} \pi(x,t|y),
\end{equation}
with the generator 
\numparts  
\begin{eqnarray} 
\label{eq:Lsa}
\hat{\mathcal{L}}_{\rm s} &=& s_0^{-1} \hat{\mathcal{L}}\, s_0 - s_0^{-1} (\hat{\mathcal{L}}\, s_0 ), \\  
&=& D \partial^2_{yy} - 
\frac{1}{\gamma}\left[ V'(y) - 2 k_{\rm B}T \frac{s_0'(y)}{s_0(y)}  \right]\! \partial_y .
\label{eq:Lsb}
\end{eqnarray}
\endnumparts
The generator~(\ref{eq:Lsa}) is obtained directly by taking the time derivative of Eq.~(\ref{eq:pi}), and using that $s_0^{-1} (\hat{\mathcal{L}}\, s_0 ) = -\lambda_0$. The explicit form~(\ref{eq:Lsb}) follows from~(\ref{eq:Lsa}) after some algebra.  
The Fokker-Planck operator adjoint to the generator $\hat{\mathcal{L}}_{\rm s}$ reads 
\numparts 
\begin{eqnarray}
\label{eq:LFPsa}
\hat{\mathcal{L}}_{\rm s}^\dagger 
&=& s_0\, \hat{\mathcal{L}}^\dagger\, s_0^{-1} -  s_0^{-1} (\hat{\mathcal{L}}\, s_0 ), \\ 
&=& D \partial^2_{xx} + 
\frac{1}{\gamma} \partial_x \left[ V'(x) - 2 k_{\rm B}T \frac{s_0'(x)}{s_0(x)}  \right]. 
\label{eq:LFPs}
\end{eqnarray}
\endnumparts 
Equivalently, the dynamics of trajectories can be described by the Langevin equation corresponding to the Fokker-Planck operator~(\ref{eq:LFPs}),
\begin{equation} 
\label{eq:langevins}
\frac{{\rm d}X_t}{{\rm d} t} = - \frac{1}{\gamma} \left[ V'(X_t) - 2 k_{\rm B}T \frac{s_0'(X_t)}{s_0(X_t)}  \right]
+ \sqrt{2 D}\, \xi_t. 
\end{equation}

In mathematics, the similarity transformation~(\ref{eq:Lsa}) is known as the Doob h-transform \cite{DoobBook}. It leaves intact the diffusion coefficient and introduces an effective drift that keeps the trajectories away from the unstable region of the potential. For a more rigorous discussion of the Q-process, we refer to the extensive mathematical literature on the topic, see e.g.\ \cite{bookQSD} and references therein. Here, our aim was to introduce these concepts heuristically and accessibly for a wide audience of physicists. 

In physics, an analogue of the Q-process can be found in theory of large deviations as one of effective processes that can be used to generate ensembles associated with rare values of time-extensive observables \cite{Jack/Sollich/PTPS2010, Chetrite/Touchette/PRL2013, Nyawo/Touchette/PRE2016, Nyawo/Touchette/PRE2018, Tizon-Escamilla/JSTAT2019, Lazarescu/etal/Arxiv2019}. For a comprehensive review of Markov processes conditioned on large deviations see Ref.~\cite{Chetrite/Touchette/AHP2015}, where also the connection to the original Doob's conditioning is described in a great detail. 

\section{Q-process: Limit distribution}
\label{sec:PIst}

For any initial condition, the dynamics of Q-process converges towards the stationary regime II with the relaxation time $1/\Delta$. The limit distribution of the Q-process can be computed as the long-time limit of its propagator, 
\begin{equation} 
\pi_{\rm st}(x) 
= \lim_{t\to\infty} \pi(x,t|y)
= \lim_{t\to\infty} \lim_{\tau \to \infty } 
\left< \delta\left( X_t - x \right) \right>_{ \tau_{\rm d}> \tau} . 
\end{equation} 
The first equality shows that any trace of the initial condition $y$ is forgotten. 
The second one emphasizes importance of correct ordering of the two limits and suggests that the limit distribution $\pi_{\rm st}(x)$ differs from the quasi-stationary distribution $Q_{\rm st}(x)$. The latter appears at the final instant $\tau$ of the transient regime III, which we have shifted to infinity before taking the limit $t\to \infty$, cf.\ Eq.~(\ref{eq:pidef}) and Eq.~(\ref{eq:Qstlimit}) below. 
  
The limit distribution $\pi_{\rm st}(x)$ describes the stationary state of trajectories which will diverge in (an infinitely) distant future. The explicit evaluation of the $t\to \infty$ limit with the aid of the leading asymptotic behavior of the propagator~(\ref{eq:propagatorasy}) yields  
\begin{equation}
\pi(x,t|y) = \frac{s_0(x) }{s_0(y)} {\rm e}^{\lambda_0 t} P(x,t|y) \sim 
\frac{s_0(x) }{s_0(y)} {\rm e}^{\lambda_0 t}\, p_0(x)s_0(y) {\rm e}^{-\lambda_0 t}, 
\quad t\to \infty .
\end{equation} 
Hence the limit distribution is given by 
\begin{equation}
\label{eq:PIstp0s0}
\pi_{\rm st}(x) = s_0(x) p_0(x).
\end{equation} 
The PDF $\pi_{\rm st}(x)$ is the stationary distribution (corresponding to the eigenvalue 0) of the Fokker-Planck operator~(\ref{eq:LFPs}). The equality 
$\hat{\mathcal{L}}^\dagger_{\rm s}\pi_{\rm st}=0$ 
can be verified directly by applying the Fokker-Planck operator~(\ref{eq:LFPsa}) on the PDF~(\ref{eq:PIstp0s0}). 

The relations~(\ref{eq:PSIp0}) and (\ref{eq:PSIs0}), between the eigenfunctions $p_0(x)$, $s_0(x)$ and the ground state wave function $\Psi_0(x)$ reveal the physical meaning of $\Psi_0(x)$:  
\begin{equation} 
\pi_{\rm st}(x) = \Psi_0^2(x).
\end{equation} 
The equality implies that powerful quantum-mechanical methods for approximating ground state wave functions can be applied to discuss $\pi_{\rm st}(x)$ for highly unstable dynamics.

Another expression for $\pi_{\rm st}(x)$ follows from the relation~(\ref{eq:PIstUsings0}) between the eigenfunctions $p_0(x)$ and $s_0(x)$ established in Sec.~\ref{sec:Schrodinger},  
\begin{equation}
\label{eq:PIsts02}
\pi_{\rm st}(x) = \frac{1}{Z}\,  {\rm e}^{-\beta V(x)} s_0^2(x) .
\end{equation} 
In this form, $\pi_{\rm st}(x)$ resembles the Gibbs canonical distribution with the density of states $g(V){\rm d} V = s_0^2(x) {\rm d}x $, and $Z=\int {\rm e}^{-\beta V} g(V) {\rm d}V$. 
The divergence of the Boltzmann factor ${\rm e}^{-\beta V}$ for $x\to -\infty$
is compensated by the fast decrease of the squared eigenfunction $s_0^2(x)$, accounting for the fact that the non-diverging trajectories do not populate the unstable region $x\ll 0$, and thus the PDF $\pi_{\rm st}(x)$ is normalizable.  

Last but not least, it is convenient to reshape Eq.~(\ref{eq:PIsts02}) into the Gibbs canonical form containing an effective potential $V_{\rm eff}$. The result is
\begin{equation} 
\label{eq:PIstVeff}
\pi_{\rm st}(x) = \frac{1}{Z}\,  {\rm e}^{- \beta V_{\rm eff}(x)}  ,
\end{equation} 
where 
\begin{equation}
\label{eq:Veff}
V_{\rm eff}(x)=V(x) + V_{\rm s}(x), 
\qquad V_{\rm s}(x)=- k_{\rm B}T \log\! \left[ s_0^2(x) \right]. 
\end{equation} 
The minus derivative of $V_{\rm s}(x)$ gives us the effective force, 
\begin{equation}
\label{eq:Fss}
F_{\rm s}(x) = 2 k_{\rm B}T \frac{s_0'(x)}{s_0(x)} ,
\end{equation}
arising from the tendency of non-diverging trajectories to avoid the highly unstable region $x\ll 0$. 
In the Langevin equation~(\ref{eq:langevins}) and in the corresponding Fokker-Planck operator~(\ref{eq:LFPs}), the force~(\ref{eq:Fss}) acts in a superposition with the actual external drift force given by $-V'(x)$. 

Even though the limit distribution of the non-diverging trajectories~(\ref{eq:PIstVeff}) and the generalized partition function~(\ref{eq:Zs2}) are formally similar to their counterparts in the equilibrium canonical ensemble, there is an important physical difference. In the equilibrium case, both $\pi_{\rm st}(x)$ and $Z$ are determined solely by the temperature and the external potential $V(x)$. 
On the contrary,  $\pi_{\rm st}(x)$ and $Z$ for the Q-process depend also on the friction constant $\gamma$ through the eigenfunction $s_0(x)$. 

Hence, $\pi_{\rm st}(x)$ and $Z$ for non-diverging trajectories are kinetic quantities whose values depend on time scales involved in the problem [and not only on the temperature and the external potential $V(x)$]. Qualitatively similar situation appears in theory of non-equilibrium steady states of processes driven by e.g.\ couplings to several reservoirs with different chemical potentials and temperatures, or/and by non-conservative external forces. In such steady states, kinetic parameters like friction coefficients (called also non-dissipative or frenetic) often determine the overall shape of the stationary probability distribution and magnitudes of the corresponding currents \cite{Basu/etal/PRL2015, Basu/etal/PCCP2015, Maes/MMCS2016, MaesNDE, Roldan/Vivo/arxiv2019}. 

Analogous strong dependencies on kinetic parameters can be expected for the limit distribution of non-diverging trajectories in highly unstable dynamics. The two cases exhibit an interesting difference in the mathematical structure of the probability distributions. In the non-equilibrium steady states, we have $s_0(x)=1$ and both the kinetic and the thermodynamic effects are mixed in a single PDF given by $p_0(x)$. On the other hand, in the highly unstable dynamics, the two effects are separated and contained within the two contributions to the effective potential~(\ref{eq:Veff}). Further investigation in this direction, focusing in particular on higher-dimensional models, may reveal new interesting phenomena for conditioned processes. 

\subsection{Tails of the limit distribution and effective potential}
\label{sec:PItails}

For large $x$ outside the unstable region, the eigenfunction $s_0(x)$ [and the potential $V_{\rm s}(x)$] approaches a constant and the force $F_{\rm s}(x)$, repelling non-diverging trajectories from the highly unstable region, vanishes. Hence the right tail of the limit distribution $\pi_{\rm st}(x)$ can be approximated by the Boltzmann tail:  
\begin{equation}
\label{eq:PIBoltzmanntail}
\pi_{\rm st}(x) \sim \frac{1}{Z} {\rm e}^{-\beta V(x)}, \qquad x\to +\infty. 
\end{equation} 
In this sense, the force $F_{\rm s}(x)$ is short-ranged and has no significant effect on dynamics for $x\gg 0$, where the effective potential in Eq.~(\ref{eq:PIstVeff}) is given by the actual external potential, $V_{\rm eff}(x) \sim V(x)$, for $x \gg 0$. 

The left tail of the limit distribution can be inferred directly from the definition of $Z$ in Eq.~(\ref{eq:Zs}) and  the eigenvalue equation for $s_0(x)$, 
\begin{equation}
- \frac{\lambda_0}{D} s_0(x) = {\rm e}^{\beta V(x)} \partial_x \left[
{\rm e}^{-\beta V(x)} s_0'(x) \right].
\label{eq:eigs02}
\end{equation}
We multiply the equation with the Boltzmann factor ${\rm e}^{-\beta V}$ and integrate over $x$. Equation~(\ref{eq:Zs}) assures that the integral is convergent and $Z$ appears on the left-hand side. After the integration, we obtain 
\begin{equation}
\label{eq:s0asylimit}
 \frac{\lambda_0}{D} Z = \lim_{x\to -\infty} {\rm e}^{-\beta V(x)} s'_0(x), 
\end{equation}
Therefore, the controlling asymptotic factor of $s_0(x)$ for $x \ll 0$ is the exponential ${\rm e}^{\beta V}$. It is straightforward to verify that the above limit is satisfied by the function   
\begin{equation}
\label{eq:s0asy}
s_0(x) \sim \frac{\lambda_0 \gamma Z}{V'(x)}\, {\rm e}^{ \beta V(x)},
\qquad x\to -\infty .
\end{equation} 
The limit distribution thus decays exponentially as  
\begin{equation}
\pi_{\rm st}(x) \sim Z \! \left[ \frac{\lambda_0 \gamma }{V'(x)} \right]^2  {\rm e}^{ \beta V(x)}, 
\qquad x\to -\infty ,
\label{eq:PIstxll0}
\end{equation} 
the exponential decay being modified by a polynomial prefactor determined by the unstable part of $V(x)$, $x\to -\infty$. Note that the constant prefactor in this equation is exact because it follows from the limit on the right-hand side of Eq.~(\ref{eq:s0asylimit}). 

The result~(\ref{eq:PIstxll0}) provides complete information on the asymptotic behavior of the total effective potential $V_{\rm eff}(x)=V(x)+V_{\rm s}(x)$ in Eq.~(\ref{eq:PIstVeff}) and in particular on its part $V_{\rm s}(x)$ arising due to the conditioning. The latter behaves as 
\begin{equation} 
\label{eq:Vasy}
V_{\rm s}(x) \sim - 2 V(x) - 2 k_{\rm B } T \log\!\left[ V'(x) \right] , 
\qquad x \to -\infty .
\end{equation}
The first term on the right-hand side compensates the external unstable potential $V(x)$ and contributes by the exponent $\beta V(x)$ responsible for the exponential decay in Eq.~(\ref{eq:s0asy}). 
The second term creates a logarithmic barrier $- 2 k_{\rm B} T (n-1) \log(x)$. 
Interestingly enough, the barrier is impenetrable for all cases of  interest in the present paper where $n>2$; for a further discussion see classification of boundary conditions for the Bessel process, e.g.\ in Refs.~\cite{Bray/PRE2000, Ryabov/etal/JCP2015}. Finally, we emphasize that the relation~(\ref{eq:Vasy}) holds only asymptotically, i.e., the effective potential has no singularity around $x=0$ where majority of non-diverging trajectories live. 

\subsection{Mean squared force}
\label{sec:Fs}

Having understood the asymptotic behaviors of the effective potential $V_{\rm eff}(x)$, it is interesting to ask, what is an average magnitude of the force $F_{\rm s}(x)$ and how it depends on the model parameters $T$, $\gamma$, $\mu$, and $n$. A simple answer is provided in this Section. It is based on the second expression for $Z$ in Eq.~(\ref{eq:Zs2}) and on a scaling analysis. 

Intuitively we expect the magnitude of the force $F_{\rm s}(x)$ to increase with the degree of instability. Indeed, it turns out that the mean squared amplitude of the force is proportional to the decay rate 
\begin{equation}
\left< F_{\rm s}^2 \right> = 4 \gamma  k_{\rm B}T \lambda_0.
\label{eq:Fs2}
\end{equation}
The average is taken with respect to $\pi_{\rm st}(x)$.  
Equation~(\ref{eq:Fs2}) follows directly from the differential equation~(\ref{eq:eigs02}) for $s_0(x)$, and from Eq.~(\ref{eq:Zs}) for the partition function containing the second power of $s_0(x)$. 
We multiply both sides of  Eq.~(\ref{eq:eigs02}) by $s_0(x) {\rm e}^{-\beta V(x)}$, integrate over $x$, identify $Z$ on the right-hand side in accordance with Eq.~(\ref{eq:Zs2}), and get
\begin{equation}
\label{eq:Fs2derivation}
\frac{\lambda_0}{D} Z = \int {\rm e}^{-\beta V(x)} s_0^2(x) \left[ \frac{s_0'(x)}{s_0(x)}\right]^{2} {\rm d} x, 
\end{equation} 
where we have performed the per partes integration on the right-hand side. After dividing Eq.~(\ref{eq:Fs2derivation}) with $Z$ and adjusting properly the $k_{\rm B}T$ factor, we get Eq.~(\ref{eq:Fs2}).  

Equation~(\ref{eq:Fs2}) is valid generally for any potential $V(x)$. Its further explicit analysis is possible for monomial unstable potentials $V(x)=\mu x^n/n$ via the scaling analysis. Defining the time and the length scales as \cite{Siler/etal/PRL2018} 
\begin{equation}
\tilde{t} = \frac{\gamma}{\mu} \left( \frac{\mu }{k_{\rm B}T} \right)^{\frac{n-2}{n}},
\qquad 
\tilde{x} = \left(\frac{k_{\rm B}T}{\mu }\right)^{\frac{1}{n}}, 
\end{equation}
respectively, the eigenvalue problem~(\ref{eq:eigenvaluepn}) attains the dimensionless form 
\begin{equation} 
\left[ \partial_{\zeta \zeta}^2 + \partial_{\zeta} \zeta^{n-1}  \right] \tilde{p}_0(\zeta) =- \theta_0 \tilde{p}_0(\zeta),
\end{equation}
where $\zeta=x/\tilde{x}$ stands for  the transformed coordinate variable, and the transformed dimensionless decay rate $\theta_0=\lambda_0 \tilde{t}$, depends on the order $n$ only. Similar transformation can be applied to the adjoint eigenvalue problem for the left eigenfunctions.

Rescaling the position variable by $\tilde{x}$ and the decay rate with $\tilde{t}$, we obtain from the exact equation~(\ref{eq:Fs2}) the explicit dependence 
\begin{equation}
\sqrt{ \left< F_{\rm s}^2 \right>} = 2 \sqrt{\theta_0} \,   \mu^{\frac{1}{n}} 
 \left( {k_{\rm B}T } \right)^{\frac{n-1}{n}} 
\end{equation} 
of the force magnitude on the model parameters.  
Similar dependence holds for the mean value of the force,
\begin{equation}
\label{eq:Fsmean}
\left< F_{\rm s} \right> = A(n) \mu^{\frac{1}{n}} 
 \left( {k_{\rm B}T } \right)^{\frac{n-1}{n}} ,
\end{equation}
where the prefactor $A$ depends on $n$ only. 

The effective force always increases with the temperature $T$ and with the potential amplitude $\mu$. The dependence on $\mu$ becomes weaker for larger powers $n$, which suppress the potential magnitude at the plateau around $x=0$, where majority of non-diverging  trajectories live. For large $n$, the temperature dependence in~(\ref{eq:Fsmean}) becomes linear for the same reason. 

\section{Quasi-stationary distribution}
\label{sec:QSD}

In this section, we compare the limit distribution of the Q-process $\pi_{\rm st}(x)$ and the quasi-stationary distribution $Q_{\rm st}(x)$. The latter arises at the end of the transient regime III in Fig.~\ref{fig:FIG1}(c). Specifically, we shall use the derived properties of the effective potential and relations between the eigenvectors to illustrate the most striking difference between the two distributions.

The statistics of trajectories at time $\tau$ that are non-diverging at least up to time $\tau$ is characterized by the conditional PDF 
\begin{equation}
\label{eq:Qdef}
Q(x,\tau|y) = \left< \delta\left( X_\tau - x \right) \right>_{ \tau_{\rm d} > \tau} = \frac{P(x,\tau|y)}{S(\tau|y)}, 
\end{equation}
where the inequality $\tau_{\rm d}>\tau$ means that each trajectory in the conditioned ensemble diverges later than at $\tau$. In contrast to Eq.~(\ref{eq:pidef}), here, the time of conditioning $\tau$ is identical with the time of observation of the process. 
Existence of the spectral gap $\Delta$ implies that the conditioned PDF~(\ref{eq:Qdef}) converges exponentially fast to the limit 
\begin{equation}
\label{eq:Qstlimit}
Q_{\rm st}(x) = \lim_{\tau \to \infty} \frac{P(x,\tau|y)}{S(\tau|y)}.
\end{equation}
The limit is known as the quasi-stationary distribution in mathematical literature, where it has first been introduced in context of branching processes and absorbed Marov chains \cite{Yaglom/Doklady1947, Mandl/CZMJ1961, Darroch/Seneta/JAP1965, Seneta/Vere-Jones/JAP1966} and since then has been studied extensively in context of population dynamics with extinction \cite{bookQSD}. The extinction event in population dynamics is in our context represented by the trajectory divergence. 

Evaluation of the limit~(\ref{eq:Qstlimit}) with the aid of the long-time expressions ~(\ref{eq:propagatorasy}) and~(\ref{eq:Sasy}) for $P(x,\tau|y)$ and $S(\tau|y)$, respectively, shows that it is equal to the eigenfunction $p_0(x)$, i.e.,  
\begin{equation}
Q_{\rm st}(x) = p_0(x). 
\end{equation}
Then, Eq.~(\ref{eq:p0usings0}) can be used to recast this relation into the form
\begin{equation}
\label{eq:Qsts0}
Q_{\rm st}(x) = \frac{1}{Z}{\rm e}^{-\beta V(x)}s_0(x),
\end{equation}
analogous to the one we have examined for $\pi_{\rm st}(x)$, cf.\ Eq.~(\ref{eq:PIsts02}). 
The two equations differ just by a power of $s_0(x)$ on their right-hand sides. The generalized partition functions $Z$ are equal in the both equations because of the identity~(\ref{eq:identityZ}), expressing that areas enclosed by nominators in Eqs.~(eq:Qsts0) and (\ref{eq:PIsts02}) are the same even though the two functions are rather different, as illustrated in the right panel of Fig.~\ref{fig:psi}.

To pursue the comparison further, the quasi-stationary distribution~(\ref{eq:Qsts0}) can be written in the form   
\begin{equation}
\label{eq:Qsteff}
Q_{\rm st}(x) =  \frac{1}{Z} {\rm e}^{-\beta [V(x) + V_{\rm s}(x)/2 ] } 
\end{equation}
resembling the canonical-like expression for $\pi_{\rm st}(x)$ in Eq.~(\ref{eq:PIstVeff}).
There is just a minor difference between expressions for the two distributions given by the factor $1/2$ in the above exponent, i.e., 
$Q_{\rm st}(x) =  {\rm e}^{\beta V_{\rm s}(x)/2 } \pi_{\rm st}(x)$.
This minor difference has no effect on the right tail of the quasi-stationary distribution. Hence the tail is the same as that of $\pi_{\rm st}(x)$ in Eq.~(\ref{eq:PIBoltzmanntail}), 
\begin{equation}
Q_{\rm st}(x) \sim \frac{1}{Z} {\rm e}^{-\beta V(x) } ,
\qquad x\to +\infty,
\end{equation}
because of converges of $s_0(x)$ to a constant. 

Contrary to the similarity of right tails, the factor $1/2$ in Eq.~(\ref{eq:Qsteff}) strongly influences the asymptotic behavior for $x\to -\infty$. Based on Eq.~(\ref{eq:s0asy}), the left tail assumes the form   
\begin{equation} 
Q_{\rm st}(x) \sim \frac{\lambda_0 \gamma }{V'(x)} ,
\qquad x\to -\infty .
\end{equation}
Specifically, for $V(x)=\mu x^n/n$ the tail decays as the power law   
\begin{equation}
\label{eq:Qstpowerlas}
Q_{\rm st}(x) \sim \frac{\lambda_0 \gamma }{\mu }\frac{1 }{ |x|^{n-1}} ,
\qquad x\to -\infty .
\end{equation}

Hence, in contrast to $V_{\rm eff}(x)=V(x)+V_{\rm s}(x)$ in the Q-process limit distribution~(\ref{eq:PIstVeff}), the modified effective potential in the present case of the quasi-stationary distribution~(\ref{eq:Qsteff}), $[V(x)+V_{\rm s}(x)/2]$, is not strong enough to confine trajectories to the stable region. 
The power-law decay~(\ref{eq:Qstpowerlas}), arising from the asymptotically logarithmic potential in Eq.~(\ref{eq:Qsteff}),
\begin{equation}
V(x)+\frac{1}{2}
V_{\rm s}(x) \sim -  k_{\rm B} T (n-1) \log(|x|),
\qquad x \to -\infty. 
\end{equation}
allows trajectories to populate any negative $x$ exploring the whole unstable region with a slowly decaying probability. 

Note that for the cubic potential, no integer moments of the quasi-stationary distribution exist and hence the standard statistical analysis of particle position based on averages provides no information on the actual particle dynamics. An alternative approach based on local quantities (position of the PDF maximum and curvature near the maximum) has been proposed \cite{Filip/Zemanek/JOpt2016, Ornigotti/etal/PRE2018} and its utility verified experimentally in \cite{Siler/etal/PRL2018} on colloidal particles diffusing in a nonlinear optical trap illustrated in Fig.~\ref{fig:FIG1}(a).  

We also note that it is possible to construct an effective conservative Markov process that converges towards the distribution $Q_{\rm st}(x)$ in the long-time limit \cite{Ornigotti/etal/PRE2018}. The process is based on an appropriate return (or resetting) of the diverging trajectories. The effective process should not be confused with the Q-process discussed above. The two has different limiting distributions and the Q-process describes faithfully the conditioned dynamics whereas the effective process of Ref.~\cite{Ornigotti/etal/PRE2018} converges towards $Q_{\rm st}(x)$ only asymptotically in the long-time limit and its transient dynamics may differ from the actual one. 

\section{Conclusions and perspectives}

We have discussed properties of the position statistics of non-diverging stochastic trajectories diffusing in highly unstable potentials. Two stationary distributions arising in this case can be formally expressed in the Gibbs canonical form with the effective potentials depending on the left eigenvector of the Fokker-Planck operator. 

The limit distribution of the Q-process describes stationary statistics of trajectories conditioned to be non-diverging for an infinitely long time. For all highly unstable systems, the distribution is localized with exponentially decaying tails. The effective force arising from the conditioning keeps trajectories away from the unstable region of the potential and its strength increases with the degree of instability. 

On the other hand, the quasi-stationary distribution is heavy-tailed towards the instability and hence it allows trajectories to be located anywhere in the unstable region. The slow power-law decay of the quasi-stationary distribution is determined by the derivative of the external potential, cf.~Eq.~(\ref{eq:Qstpowerlas}). Here, an interesting question arises whether the theory of non-normalizable distributions \cite{Barkai/PRL2019} may contribute to our understanding of unstable dynamics with non-integer exponents $n$ smaller than 3. 

As shown in Sec.~\ref{sec:Schrodinger}, the basic property of highly unstable potentials is that the spectrum of the Fokker-Planck operator is negative and discrete. Other systems satisfying the same property include confined diffusions with absorbing boundaries. At the same time, our results  imply that the magnitude of the effective force, which confines conditioned trajectories of the Q-process, decreases with the decreasing instability of the potential, cf.~Sec.~\ref{sec:Fs}. Therefore, it may be interesting to explore the dynamics for other unstable potentials and identify the precise conditions, when the Q-process becomes transient. 

Finally, we remark that the similarity of the two distributions with the Gibbs canonical ensemble is purely formal. In general, the distributions for the highly unstable systems, depended on kinetic parameters such as the friction coefficients. This dependence implies a remarkable possibility to control the overall shape of the two distributions by varying the kinetic parameters only. This effect should become more pronounced in higher-dimensional systems and it represents another interesting area for a further investigation. 

\ack 
We gratefully acknowledges financial support by the Czech Science Foundation (project No. 17-06716S). VH also thanks for support by the Humboldt foundation. AR is grateful to organizers of the {\em nesmcq18} conference, where a part of the present work has been discussed.  AR acknowledges financial support from the Portuguese Foundation for Science and Technology (FCT) under Contracts nos. PTDC/FIS-MAC/28146/2017 (LISBOA-01-0145-FEDER-028146) and UID/FIS/00618/2019.

\section*{References}
\bibliography{referencesQP}

\providecommand{\newblock}{}
\begin{thebibliography}{10}
\expandafter\ifx\csname url\endcsname\relax
  \def\url#1{{\tt #1}}\fi
\expandafter\ifx\csname urlprefix\endcsname\relax\def\urlprefix{URL }\fi
\providecommand{\eprint}[2][]{\url{#2}}

\bibitem{Siler/etal/SciRep2017}
{\v S}iler M, J{\' a}kl P, Brzobohatý O, Ryabov A, Filip R and Zem{\' a}nek P
  2017 {\em Sci. Rep.\/} {\bf 7} 1697
  \urlprefix\url{https://doi.org/10.1038/s41598-017-01848-4}

\bibitem{Siler/etal/PRL2018}
\ifmmode~\check{S}\else \v{S}\fi{}iler M, Ornigotti L, Brzobohat\'y O, J\'akl
  P, Ryabov A, Holubec V, Zem\'anek P and Filip R 2018 {\em Phys. Rev. Lett.\/}
  {\bf 121} 230601
  \urlprefix\url{https://doi.org/10.1103/PhysRevLett.121.230601}

\bibitem{Filip/Zemanek/JOpt2016}
Filip R and Zem{\' a}nek P 2016 {\em J. Opt.\/} {\bf 18} 065401
  \urlprefix\url{https://doi.org/10.1088/2040-8978/18/6/065401}

\bibitem{Zemanek/etal/JOpt2016}
Zem{\' a}nek P, {\v S}iler M, Brzobohat{\' y} O, J{\' a}kl P and Filip R 2016
  {\em J. Opt.\/} {\bf 18} 065402
  \urlprefix\url{https://doi.org/10.1088/2040-8978/18/6/065402}

\bibitem{Ornigotti/etal/PRE2018}
Ornigotti L, Ryabov A, Holubec V and Filip R 2018 {\em Phys. Rev. E\/} {\bf 97}
  032127 \urlprefix\url{http://doi.org/10.1103/PhysRevE.97.032127}

\bibitem{Chetrite/Touchette/AHP2015}
Chetrite R and Touchette H 2015 {\em Ann. Henri Poincar{\'e}\/} {\bf 16} 2005
  \urlprefix\url{https://doi.org/10.1007/s00023-014-0375-8}

\bibitem{bookQSD}
Collet P, Mart\'inez S and {San Mart\'in} J 2013 {\em Quasi-Stationary
  Distributions: Markov Chains, Diffusions and Dynamical Systems\/}
  (Springer-Verlag Berlin Heidelberg) ISBN 978-3-642-33130-5
  \urlprefix\url{https://doi.org/10.1007/978-3-642-33131-2}

\bibitem{DoobBook}
Doob J~L 2001 {\em Classical Potential Theory and Its Probabilistic
  Counterpart\/} (Springer-Verlag Berlin Heidelberg) ISBN 978-3-540-41206-9
  \urlprefix\url{https://doi.org/10.1007/978-3-642-56573-1}

\bibitem{RednerBook}
Redner S 2001 {\em {A Guide to First-Passage Processes}\/} (Cambridge
  University Press) ISBN 9780511606014
  \urlprefix\url{https://doi.org/10.1017/CBO9780511606014}

\bibitem{HanggiRevModPhys1990}
H{\" a}nggi P, Talkner P and Borkovec M 1990 {\em Rev. Mod. Phys.\/} {\bf 62}
  251--341 \urlprefix\url{https://doi.org/10.1103/RevModPhys.62.251}

\bibitem{Arecchi1982}
Arecchi F~T, Politi A and Ulivi L 1982 {\em Il Nuovo Cimento B (1971-1996)\/}
  {\bf 71} 119--154 \urlprefix\url{http://dx.doi.org/10.1007/BF02721698}

\bibitem{YoungPRA1985}
Young M~R and Singh S 1985 {\em Phys. Rev. A\/} {\bf 31}(2) 888--891
  \urlprefix\url{https://doi.org/10.1103/PhysRevA.31.888}

\bibitem{SanchoPRA1989}
Colet P, San~Miguel M, Casademunt J and Sancho J~M 1989 {\em Phys. Rev. A\/}
  {\bf 39} 149--156 \urlprefix\url{https://doi.org/10.1103/PhysRevA.39.149}

\bibitem{HirschPRA1982}
Hirsch J~E, Huberman B~A and Scalapino D~J 1982 {\em Phys. Rev. A\/} {\bf 25}
  519--532 \urlprefix\url{https://doi.org/10.1103/PhysRevA.25.519}

\bibitem{Sigeti1989}
Sigeti D and Horsthemke W 1989 {\em J. Stat. Phys.\/} {\bf 54} 1217--1222
  \urlprefix\url{http://dx.doi.org/10.1007/BF01044713}

\bibitem{SanchoPRA1991}
Ram{\' i}rez-Piscina L and Sancho J~M 1991 {\em Phys. Rev. A\/} {\bf 43}
  663--668 \urlprefix\url{https://doi.org/10.1103/PhysRevA.43.663}

\bibitem{Caceres1995}
C{\' a}ceres M~O, Budde C~E and Sibona G~J 1995 {\em J. Phys. A: Math. Gen.\/}
  {\bf 28} 3877 \urlprefix\url{https://doi.org/10.1088/0305-4470/28/14/009}

\bibitem{MantegnaPRL1996}
Mantegna R~N and Spagnolo B 1996 {\em Phys. Rev. Lett.\/} {\bf 76} 563--566
  \urlprefix\url{https://doi.org/10.1103/PhysRevLett.76.563}

\bibitem{AgudovPRE1998}
Agudov N~V 1998 {\em Phys. Rev. E\/} {\bf 57} 2618--2625
  \urlprefix\url{https://doi.org/10.1103/PhysRevE.57.2618}

\bibitem{Lindner2003}
Lindner B, Longtin A and Bulsara A 2003 {\em Neural Comput.\/} {\bf 15}
  1761--1788 \urlprefix\url{https://doi.org/10.1162/08997660360675035}

\bibitem{Brunel2003}
Brunel N and Latham P~E 2003 {\em Neural Comput.\/} {\bf 15} 2281--2306
  \urlprefix\url{https://doi.org/10.1162/089976603322362365}

\bibitem{FiasconaroPRE2005}
Fiasconaro A, Spagnolo B and Boccaletti S 2005 {\em Phys. Rev. E\/} {\bf 72}
  061110 \urlprefix\url{https://doi.org/10.1103/PhysRevE.72.061110}

\bibitem{CaceresJSP2008}
C{\'a}ceres M~O 2008 {\em J. Stat. Phys.\/} {\bf 132} 487--500
  \urlprefix\url{http://dx.doi.org/10.1007/s10955-008-9554-7}

\bibitem{Ryabov/etal/PRE2016}
Ryabov A, Zem\'anek P and Filip R 2016 {\em Phys. Rev. E\/} {\bf 94} 042108
  \urlprefix\url{http://doi.org/10.1103/PhysRevE.94.042108}

\bibitem{AgudovMalakhovPRE1999}
Agudov N~V and Malakhov A~N 1999 {\em Phys. Rev. E\/} {\bf 60} 6333--6342
  \urlprefix\url{https://doi.org/10.1103/PhysRevE.60.6333}

\bibitem{DubkovPRE2004}
Dubkov A~A, Agudov N~V and Spagnolo B 2004 {\em Phys. Rev. E\/} {\bf 69} 061103

\bibitem{GardinerSM}
Gardiner C 2009 {\em {Stochastic Methods: A Handbook for the Natural and Social
  Sciences}\/} 4th ed (Springer, Berlin, Heidelberg) ISBN 978-3-540-70712-7
  \urlprefix\url{http://www.springer.com/978-3-540-70712-7}

\bibitem{HakenSynergetics}
Haken H 2004 {\em {Synergetics: Introduction and Advanced Topics}\/} (Springer,
  Berlin, Heidelberg) ISBN 978-3-642-07405-9
  \urlprefix\url{https://doi.org/10.1007/978-3-662-10184-1}

\bibitem{RiskenFPE}
Risken H 1996 {\em {The Fokker-Planck Equation: Methods of Solution and
  Applications}\/} 2nd ed (Springer, Berlin, Heidelberg) ISBN 978-3-540-61530-9
  \urlprefix\url{https://doi.org/10.1007/978-3-642-61544-3}

\bibitem{VanKampenSP}
{Van Kampen} N~G 2007 {\em {Stochastic Processes in Physics and Chemistry}\/}
  3rd ed (North Holland) ISBN 978-0-444-52965-7
  \urlprefix\url{https://doi.org/10.1016/B978-0-444-52965-7.X5000-4}

\bibitem{Berezin/Shubin/SE}
Berezin F~A and Shubin M 1991 {\em The Schr\"odinger Equation\/} (Springer
  Netherlands) ISBN 978-0-7923-1218-5
  \urlprefix\url{https://doi.org/10.1007/978-94-011-3154-4}

\bibitem{Lambert/EJP2007}
Lambert A 2007 {\em Electron. J. Probab.\/} {\bf 12} 420
  \urlprefix\url{https://doi.org/10.1214/EJP.v12-402}

\bibitem{Meleard/Villemonais/PS2012}
M\'el\'eard S and Villemonais D 2012 {\em Probab. Surveys\/} {\bf 9} 340
  \urlprefix\url{https://doi.org/10.1214/11-PS191}

\bibitem{Jack/Sollich/PTPS2010}
Jack R~L and Sollich P 2010 {\em Prog. Theor. Phys. Suppl.\/} {\bf 184} 304
  \urlprefix\url{https://doi.org/10.1143/PTPS.184.304}

\bibitem{Chetrite/Touchette/PRL2013}
Chetrite R and Touchette H 2013 {\em Phys. Rev. Lett.\/} {\bf 111} 120601
  \urlprefix\url{https://doi.org/10.1103/PhysRevLett.111.120601}

\bibitem{Nyawo/Touchette/PRE2016}
Nyawo P~T and Touchette H 2016 {\em Phys. Rev. E\/} {\bf 94} 032101
  \urlprefix\url{https://doi.org/10.1103/PhysRevE.94.032101}

\bibitem{Nyawo/Touchette/PRE2018}
Nyawo P~T and Touchette H 2018 {\em Phys. Rev. E\/} {\bf 98} 052103
  \urlprefix\url{https://doi.org/10.1103/PhysRevE.98.052103}

\bibitem{Tizon-Escamilla/JSTAT2019}
Tiz{\'{o}}n-Escamilla N, Lecomte V and Bertin E 2019 {\em J. Stat. Mech.\/}
  {\bf 2019} 013201 \urlprefix\url{https://doi.org/10.1088/1742-5468/aaeda3}

\bibitem{Lazarescu/etal/Arxiv2019}
{Lazarescu} A, {Cossetto} T, {Falasco} G and {Esposito} M 2019 {\em arXiv
  e-prints\/} arXiv:1902.08416 \urlprefix\url{https://arxiv.org/abs/1902.08416}

\bibitem{Basu/etal/PRL2015}
Basu U, Maes C and Neto\ifmmode~\check{c}\else \v{c}\fi{}n\'y K 2015 {\em Phys.
  Rev. Lett.\/} {\bf 114} 250601
  \urlprefix\url{https://doi.org/10.1103/PhysRevLett.114.250601}

\bibitem{Basu/etal/PCCP2015}
Basu U, Krüger M, Lazarescu A and Maes C 2015 {\em Phys. Chem. Chem. Phys.\/}
  {\bf 17} 6653 \urlprefix\url{http://dx.doi.org/10.1039/C4CP04977B}

\bibitem{Maes/MMCS2016}
Maes C 2016 {\em Math. Mech. Complex Syst.\/} {\bf 4} 275
  \urlprefix\url{http://dx.doi.org/10.2140/memocs.2016.4.275}

\bibitem{MaesNDE}
Maes C 2018 {\em {Non-Dissipative Effects in Nonequilibrium Systems}\/}
  (Springer International Publishing) ISBN 978-3-319-67779-8
  \urlprefix\url{https://doi.org/10.1007/978-3-319-67780-4}

\bibitem{Roldan/Vivo/arxiv2019}
{Rold{\'a}n} {\'E} and {Vivo} P 2019 {\em arXiv e-prints\/}  arXiv:1903.08271
  \urlprefix\url{https://arxiv.org/abs/1903.08271}

\bibitem{Bray/PRE2000}
Bray A~J 2000 {\em Phys. Rev. E\/} {\bf 62} 103
  \urlprefix\url{https://doi.org/10.1103/PhysRevE.62.103}

\bibitem{Ryabov/etal/JCP2015}
Ryabov A, Berestneva E and Holubec V 2015 {\em J. Chem. Phys.\/} {\bf 143}
  114117 \urlprefix\url{https://doi.org/10.1063/1.4931474}

\bibitem{Yaglom/Doklady1947}
Yaglom A~M 1947 {\em Dokl. Acad. Nauk SSSR (in Russian)\/} {\bf 56} 795

\bibitem{Mandl/CZMJ1961}
Mandl P 1961 {\em Czechoslovak Math. J.\/} {\bf 11} 558
  \urlprefix\url{http://eudml.org/doc/12097}

\bibitem{Darroch/Seneta/JAP1965}
Darroch J~N and Seneta E 1965 {\em J. Appl. Probab.\/} {\bf 2} 88
  \urlprefix\url{https://doi.org/10.2307/3211876}

\bibitem{Seneta/Vere-Jones/JAP1966}
Seneta E and Vere-Jones D 1966 {\em J. Appl. Probab.\/} {\bf 3} 403
  \urlprefix\url{https://doi.org/10.2307/3212128}

\bibitem{Barkai/PRL2019}
Aghion E, Kessler D~A and Barkai E 2019 {\em Phys. Rev. Lett.\/} {\bf 122}
  010601 \urlprefix\url{https://doi.org/10.1103/PhysRevLett.122.010601}

\end{thebibliography}

\end{document}